\documentclass[aip,reprint,amsmath,amssymb,floatfix]{revtex4-1}

\usepackage{graphicx}

\usepackage{color}

\usepackage{chemfig}

\newcommand{\BE}{\begin{equation}}
\newcommand{\EE}{\end{equation}}

\newcommand{\BEA}{\begin{eqnarray}}
\newcommand{\EEA}{\end{eqnarray}}

\newcommand{\Sec}[1]{Sec.~\ref{sec:#1}}
\newcommand{\Eq}[1]{Eq.~\ref{eq:#1}}
\newcommand{\Fig}[1]{Fig.~\ref{fig:#1}}
\newcommand{\Table}[1]{Table~\ref{tab:#1}}

\renewcommand{\Ref}[1]{Ref.~\cite{#1}}  

\newcommand{\Lp}{L_p}
\newcommand{\Lpmax}{L_p^\mathrm{max}}

\newcommand{\tfull}{t_\mathrm{full}}

\newcommand{\Tmax}{T_\mathrm{max}}
\newcommand{\Nrep}{N_\mathrm{rep}}

\newcommand{\trep}{t_\mathrm{rep}}
\newcommand{\ttot}{t_\mathrm{tot}}

\newcommand{\Nmon}{N_\mathrm{mon}}
\newcommand{\Nsolv}{N_\mathrm{solv}}

\newcommand{\Lbox}{L_\mathrm{box}}

\newcommand{\Euu}{E_\mathrm{uu}}
\newcommand{\Euv}{E_\mathrm{uv}}
\newcommand{\Evv}{E_\mathrm{vv}}

\newcommand{\Escore}{E_\mathrm{sc}}
\newcommand{\Eint}{E_\mathrm{int}}

\newcommand{\Gtot}{G_\mathrm{tot}}
\newcommand{\Gsolv}{G_\mathrm{solv}}

\begin{document}

\title{Sequential replica exchange with solute tempering for atomistic modeling of supramolecular polymer structures}



\author{Hadi H. Arefi}
\email{hadi.arefi.8u@kyoto-u.jp}

\author{Takeshi Yamamoto}

\affiliation{Department of Chemistry, Graduate School of Science, Kyoto University, Kyoto 606-8502, Japan}
\email{yamamoto@kuchem.kyoto-u.ac.jp}

\maketitle
\date{\today}


\section*{Abstract} \label{sec:abst}  


Atomistic modeling of supramolecular polymers remains challenging
because spontaneous self-assembly and structural relaxation can be slow,
while simulations starting from preconstructed structures may retain bias
from the assumed initial architecture.
Here, we present a construction-based atomistic modeling strategy
that combines sequential monomer addition with localized enhanced sampling.
Starting from a short seed, monomers are added one at a time to the growing terminus,
and replica exchange with solute tempering (REST) is applied locally
to the incoming monomer to explore alternative binding configurations
before elongation proceeds.
We examine this strategy using benzene-1,3,5-tricarboxamide (BTA)
with hexyl side chains in explicit methylcyclohexane.
Benchmark calculations with conventional MD (cMD) and global REST reveal
increasing sampling difficulties with system size; for global REST,
these include growing replica requirements and inefficient replica traversal
associated with aggregation-dispersion behavior.
In contrast, sequential REST constructs BTA stacks up to 40 monomers
while maintaining the characteristic 2:1 hydrogen-bonding motif
observed in global REST calculations, whereas a corresponding
sequential cMD test becomes kinetically trapped and
fails to sustain ordered elongation, thus highlighting
the utility of localized enhanced sampling during stepwise growth.
The present method is not intended to reproduce spontaneous polymerization
kinetics or a complete equilibrium ensemble. Rather, it provides
a construction-based route for generating atomistic models
of one-dimensional supramolecular assemblies
while mitigating the size-dependent sampling difficulties associated with global REST.

\section{Introduction} \label{sec:intro}  

Supramolecular polymers are one-dimensional assemblies formed through
reversible noncovalent interactions between monomers.
Because these interactions are dynamic and act collectively within the assembly,
the resulting structures can be sensitive to monomer chemistry and environmental conditions.
\cite{DeGreefReview2008,DeGreefReview2009,AidaReview2012,MeijerReview2025}
Molecular simulations can provide atomistic information
on their intermolecular organization and thereby help connect
monomer-level interactions with the structure of the resulting polymer.
Obtaining reliable atomistic models of supramolecular polymers, however,
remains challenging because aggregation and structural relaxation
often occur on time scales that are difficult to access
with conventional molecular dynamics (MD).
In particular, formation of an aggregate does not necessarily
imply formation of an ordered polymer: monomers may become trapped
in metastable binding configurations that require substantial
rearrangement before a well-ordered structure is obtained.
\cite{PavanReview2018,BalaReview2018,MarrinkReview2018}

Two general strategies are commonly used to obtain molecular models
of supramolecular polymers.\cite{PavanReview2018,Pavan2016water,vanEsch2021}
In top-down modeling, a plausible aggregate structure is constructed
in advance and subsequently relaxed under the conditions of interest.
This approach makes atomistic modeling of relatively large and complex
assemblies feasible, but its reliability depends on whether
the preconstructed structure can sufficiently reorganize
within the accessible simulation time.
In bottom-up simulations, by contrast, initially dispersed monomers
undergo aggregation and structural organization during the simulation.
This reduces the need to prescribe the final polymer structure,
but atomistic bottom-up assembly can be hindered by kinetic trapping,
the large configurational space of dispersed monomers,
and the slow correction of defective binding configurations.
Thus, the two approaches involve a trade-off between
the structural assumptions introduced during model construction
and the amount of configurational sampling required to obtain an ordered assembly.

Enhanced-sampling methods have been used to mitigate sampling limitations
in both types of modeling.\cite{PavanReview2018,BalaReview2018,MarrinkReview2018}
For top-down models, Bochicchio and Pavan applied accelerated MD and metadynamics
to a preconstructed atomistic BTA supramolecular fiber in water and
showed that enhanced sampling can accelerate its structural relaxation
and provide more reproducible equilibrated configurations than
conventional MD.\cite{Pavan2018arxiv}
In this strategy, the complete polymer is constructed beforehand
and enhanced sampling is used to relax the preassembled structure.
In a complementary bottom-up setting, our previous study applied
replica exchange with solute tempering (REST) to initially dispersed
BTA monomers in explicit methylcyclohexane solvent.\cite{Arefi2017}
REST generated ordered columnar structures with characteristic
helical hydrogen-bonding patterns that were not obtained by conventional MD,
demonstrating that enhanced sampling can facilitate both aggregation
and error correction at the atomistic level.

Nevertheless, global enhanced sampling based on replica-exchange methods
becomes increasingly demanding as the size of the assembling system increases.
In both temperature-based replica-exchange MD (REMD) and REST,
sufficiently strong tempering is required to rearrange
misbound monomers, whereas strong tempering can also induce monomer
dissociation and an aggregation--dispersion transition
that slows replica traversal.
More generally, when all monomers in a growing aggregate participate
in enhanced sampling, the configurational problem
to be explored increases together with the target polymer size.

Here, we pursue a different strategy in which polymer construction
and enhanced sampling are interleaved during growth.
Assuming a predominantly one-dimensional morphology,
we construct the aggregate sequentially rather than introducing
all monomers at once or preconstructing the complete target polymer.
At each step, an incoming monomer is placed near the terminus
of an existing aggregate and subjected to local REST sampling
to explore alternative binding configurations.
A favorable configuration is then selected, the monomer is incorporated
into the growing aggregate, and the procedure is repeated.

The resulting strategy occupies an intermediate position
between conventional top-down and fully bottom-up modeling:
the complete polymer structure is not specified in advance,
while spontaneous assembly and global rearrangement
of all monomers are not sampled simultaneously.
It also differs from enhanced relaxation of a preconstructed fiber,
because sampling is applied locally as the structure is being built
rather than globally after construction of the complete assembly.
In this way, a single growing global sampling problem is replaced
by a sequence of smaller local sampling problems.
%
The sequential strategy also defines the limitations of the method.
The resulting structure can depend on the growth history
and on the configurations selected at preceding steps,
and large-scale rearrangements of previously incorporated monomers
are not sampled exhaustively. The method is therefore intended
as an atomistic structure-modeling strategy under a one-dimensional growth
assumption rather than as a procedure for obtaining the equilibrium
ensemble of the complete polymer.

In this work, we examine this sequential strategy using an all-atom model
of a BTA-based supramolecular polymer in explicit methylcyclohexane solvent.
We first investigate how global REST sampling changes with increasing
numbers of monomers and identify limitations relevant to the construction
of longer polymers. We then apply sequential monomer addition
combined with local REST sampling at the polymer terminus
and examine whether favorable binding configurations
and characteristic intermolecular hydrogen-bonding motifs
can be obtained while restricting enhanced sampling
to the growing region of the aggregate.
By comparing the resulting structures with global REST calculations,
we assess the applicability and limitations of the sequential approach.
The results suggest a practical route for constructing atomistically
resolved models of one-dimensional supramolecular polymers
when fully global enhanced sampling becomes increasingly demanding.



\begin{figure}[tb]
\centering
\setchemfig{atom sep=2.2em, bond style={line width=0.8pt}, atom style={font=\small}}
\chemfig{*6((-(=[::60]O)-[::-60]N(-[::-60]H)-C_6H_{13})=(-H)-(-(=[::60]O)-[::-60] N(-[::-60]H)-C_6H_{13})=(-H)-(-(=[::60]O)-[::-60]N(-[::-60]H)-C_6H_{13})=(-H)-)}
\caption{\label{fig:monomer}
Structure of benzene-1,3,5-tricarboxamide (BTA) molecule with hexyl side chains
[-(CH$_2$)$_5$CH$_3$], denoted hereafter as BTA-C6.
}
\end{figure}


\begin{figure}[tb]
  \centering
  \includegraphics[width=1.0\columnwidth]{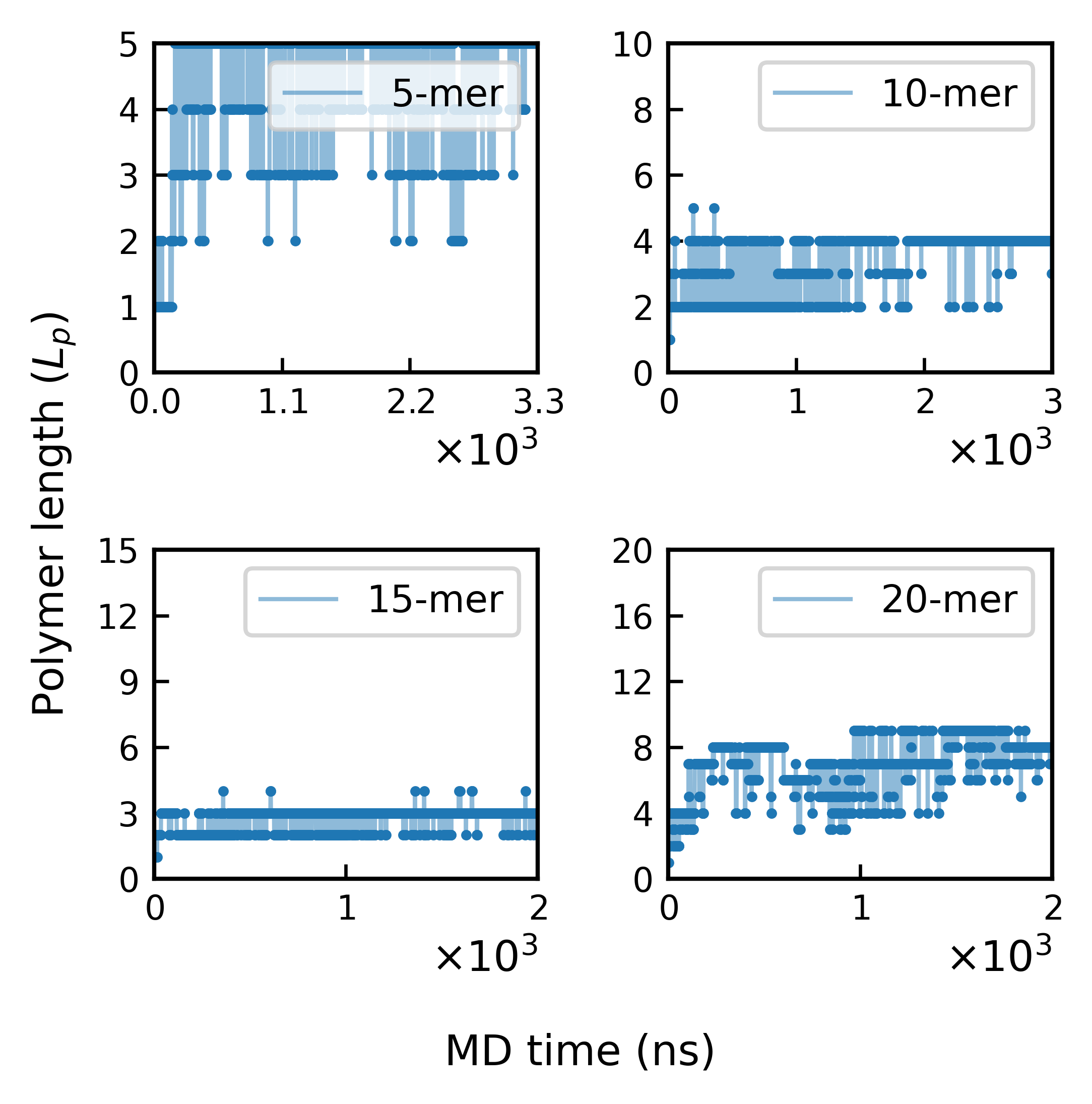}
  \includegraphics[width=0.9\columnwidth]{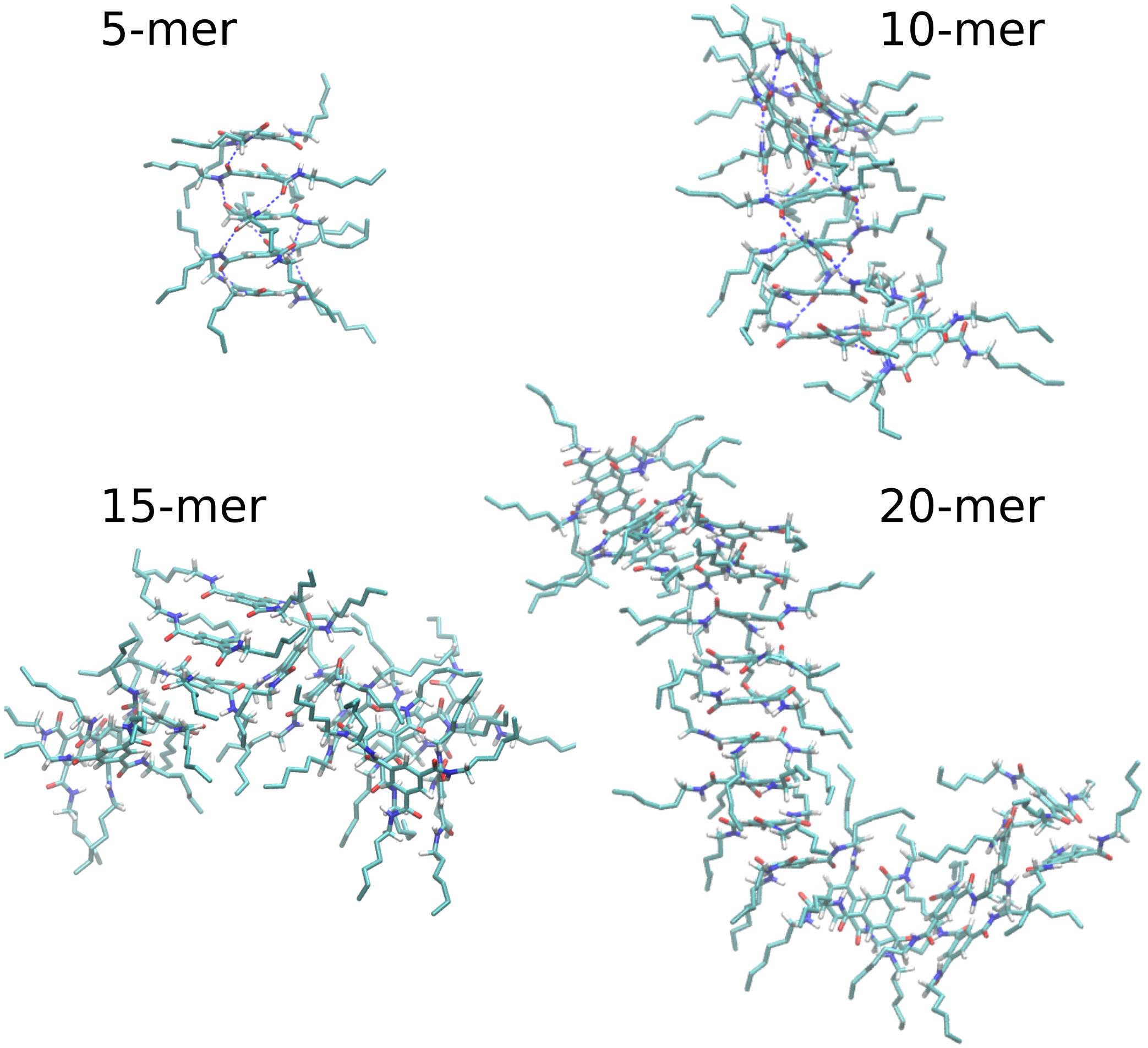}
\caption{\label{fig:cmd_polylen}
Evolution of the polymer length $L_p(t)$
obtained from cMD simulations for systems with 5, 10, 15, and 20 monomers.
Typical self-assembled structures are also shown.
cMD fails to efficiently produce elongated polymers
because of kinetically trapped structures with improperly formed H-bonds.
}
\end{figure}


\begin{table}[tb]
\caption{\label{tab:cmd}
Results of cMD simulations for systems of $\Nmon$ BTA and $\Nsolv$ MCH molecules
contained in a cubic box of side length $\Lbox$.
$t_\mathrm{MD}$ is the duration of each simulation.
$\Lpmax$ is the maximum value of $L_p$ (the length of an elongated polymer).
The monomer concentration is fixed at 0.038 M (0.023 molecules/nm$^3$)
for all the systems.
}
\begin{ruledtabular}
\begin{tabular}{ccccc}
$\Nmon$  &  $\Nsolv$  &  $\Lbox$(\AA)
&  $t_\mathrm{MD}$(ns) &  $\Lpmax$  \\
\hline
 5  &  1000  &  60.1  &  3000  &  5   \\
10  &  2000  &  75.8  &  3000  &  5   \\
15  &  3000  &  86.8  &  2000  &  4   \\
20  &  4000  &  95.4  &  2000  &  9   \\
\end{tabular}
\end{ruledtabular}
\end{table}


\section{Conventional MD} \label{sec:cmd}  


In our previous study,\cite{Arefi2017} we examined the applicability of REST
for relatively small systems and demonstrated the formation of
elongated polymers in fully bottom-up simulations.
Building on that work, the present and following sections
systematically examine the scalability and limitations
of conventional MD and global REST for increasing system sizes.


As a prototypical test system, we consider an all-atom model of
benzene-1,3,5-tricarboxamide (BTA) molecules dissolved in
explicit methylcyclohexane (MCH) solvent. BTA consists of
a benzene core, three amide groups, and alkyl side chains (\Fig{monomer}).
The amide groups can form a three-fold H-bonding network with neighboring monomers,
providing a major driving force for one-dimensional columnar assembly.
Different alkyl side chains give rise to aggregates with diverse morphologies
and properties.\cite{Cantekin2012}
Here we consider BTA molecules with hexyl side chains
(hereafter referred to as BTA-C6)
to facilitate comparison with earlier studies.\cite{Bejagam2014,Arefi2017}

As a reference, we first tested the performance of conventional MD (cMD).
Several simulations were performed for different system sizes,
with the monomer concentration fixed at 0.038 M for all systems (\Table{cmd}).
To study fully bottom-up assembly, the simulations were initialized
from molecularly dispersed states in which monomers were
randomly distributed throughout the solvent.
Solute and solvent molecules were modeled using
the GAFF and OPLS-UA force fields,
respectively. Further details of the simulation protocols are provided
in the Supplementary Material.



To characterize the polymerization of the system,
we consider two order parameters,\cite{Arefi2017}
$N_c$ and $L_p$. The $N_c(t)$ represents the number of independent
clusters observed in the system at time $t$.
Here, an independent cluster refers to a set of monomers with
all the nearest-neighbor distances less than a given
threshold (here 8 \AA{}), with the distance measured using the center of
benzene cores. With the above definition
$N_c$ = 1 corresponds to a fully aggregated state,
whereas $N_c$ = $\Nmon$ indicates a molecularly dispersed state
(where $\Nmon$ is the total number of monomers).

On the other hand, $L_p(t)$ represents the maximum length of
neatly stacked one-dimensional BTA polymers observed in the system.
Here a set of monomers are regarded as neatly stacked
when all the nearest-neighbor distances are less than 4 \AA{}
(which occurs only when the benzene cores are tightly stacked to each other).
With this definition $L_p = \Nmon$ means that all the monomers
form a single, fully elongated polymer, whereas $L_p$ = 1 means that
the system contains no stacked BTAs and thus represents a disordered state.
We note that the same definitions of $N_c$ and $L_p$ were used
in our previous study.\cite{Arefi2017}


We first examined the aggregation behavior starting from the initial dispersed states.
All systems rapidly formed a single aggregate ($N_c = 1$) within $\sim$200 ns,
except for the 20-mer system, which required a longer aggregation time ($\sim$1 $\mu$s).
\Fig{cmd_polylen} displays the evolution of $L_p$, which monitors
the formation of supramolecular polymers. Despite the rapid aggregation,
elongated polymers with $L_p \ge 10$ were not produced
within the accessible simulation time.
The limited growth of $L_p$ originates from kinetic trapping
in metastable states with improperly formed H-bonds.
The trajectory shows that short polymer fragments
tend to associate with each other in a T-shaped or kinked manner.
Once such structures are formed, further rearrangement toward
more ordered states becomes infrequent within the accessible simulation time scale.
Such kinetically trapped states are also reported
in previous studies\cite{Pavan2016water,Arefi2017,vanEsch2021}
and imply the limited sampling capability of cMD.


\begin{table}[tb]
\caption{\label{tab:grest}
Results of global REST simulations for systems with 10, 20, and 30 BTA monomers
dissolved in 2000, 4000, and 8060 MCH molecules, respectively.
$\trep$ is the MD time per replica,
and $\ttot$ is the total MD time defined by $\ttot = \trep \times \Nrep$.
$\Lpmax$ is the maximum value of $L_p(t)$ obtained for the target replica at 300 K.
The simulation box sizes for the 10, 20, and 30-mer systems are 76, 95, and 120 \AA{},
respectively, corresponding to monomer concentrations of 0.038, 0.038, and 0.028 M.
}
\begin{ruledtabular}
\begin{tabular}{ccccc}
$\Tmax(K)$  &  $\Nrep$  &  $\trep$(ns)  &  $\ttot$(ns)  &  $\Lpmax$  \\
\hline
\multicolumn{5}{c}{10-mer} \\
382  &   4  &  600  &  2400  &  10 \\
446  &   6  &  400  &  2400  &  10 \\
518  &   8  &  300  &  2400  &  10 \\
600  &  10  &  240  &  2400  &  10 \\
\hline
\multicolumn{5}{c}{20-mer} \\
370  &   5  &  500  &  2500  &  13  \\ 
390  &   6  &  500  &  3000  &  15  \\ 
460  &  12  &  500  &  6000  &  20  \\ 
\hline
\multicolumn{5}{c}{30-mer} \\
400  &  12  & 1500  & 18000  & 12 \\ 
\end{tabular}
\end{ruledtabular}
\end{table}


\section{Global REST} \label{sec:grest}  

\subsection{Replica exchange with solute tempering}

REST\cite{Liu2005,Huang2007,Wang2011}
is a Hamiltonian replica exchange method\cite{Sugita2000}
that accelerates sampling of a selected ``solute'' region
via effective high-temperature replicas,
while evolving the remaining ``solvent'' region at ambient temperature.
Specifically, we use the modified energy function for replica
$m = 0, \ldots, \Nrep - 1$
based on the REST2 scheme:\cite{Terakawa2010,Wang2011,Bussi2013}
\BE
\label{eq:rest}
  E_m =
    \frac{ \beta_m }{ \beta_0 } \Euu
  + \sqrt{ \frac{ \beta_m }{ \beta_0 } } \Euv
  + \Evv
  ,
\EE
where $\Euu$ is the solute energy, $\Euv$ the solute-solvent interaction energy,
and $\Evv$ the bulk solvent energy.
$\beta_m = 1/(k_B T_m)$ is the effective reciprocal temperature for replica $m$.
We emphasize that all replicas in REST are simulated at the target temperature
$T_0$ using the effective energy function above.
In this method, the exchange probability is independent of $\Evv$,
thereby significantly reducing the number of required replicas
compared to T-REMD.\cite{Wang2011}

In the following we study the performance of REST for the same BTA-C6 system.
Here all BTA molecules are treated as the ``solute'' region
and all MCH molecules as the ``solvent'' region.
We refer to this type of simulation as global REST
because all BTA molecules are accelerated uniformly.
The key simulation parameters
are the minimum and maximum replica temperatures ($T_0$ and $\Tmax$)
and the number of replicas ($\Nrep$) spanning the temperature range.
In this paper we set the minimum replica temperature $T_0$ at 300 K
and determined the other temperatures using a geometric rule:
\BE
  T_m = T_0 \left( \frac{ \Tmax }{ T_0 } \right)^{m/(\Nrep - 1)}
  .
\EE
We note that the present simulation protocol is essentially identical
to that used in our previous study,\cite{Arefi2017}
except that $\Tmax$ and $\Nrep$ are chosen separately for each system.

\subsection{Global REST for the 10-mer system}

We first examine the performance of REST for the same 10-mer system studied in \Sec{cmd}.
REST simulations were performed for several different values
of $\Tmax$ (382, 446, 518, and 600 K).
For each value of $\Tmax$, the number of replicas was set to
$\Nrep = 4$, 6, 8, and 10, respectively,
to maintain an average exchange probability of $\sim$20 \%.
The MD time per replica ($\trep$) was chosen
so that the computational cost was identical across all REST simulations.
We note that the total MD time of a single REST run
(defined by $\ttot = \Nrep \times \trep$) is 2400 ns,
which is smaller than that of the corresponding cMD simulation (3000 ns).

\begin{figure}[tb]
  \centering
  \includegraphics[width=0.95\columnwidth]{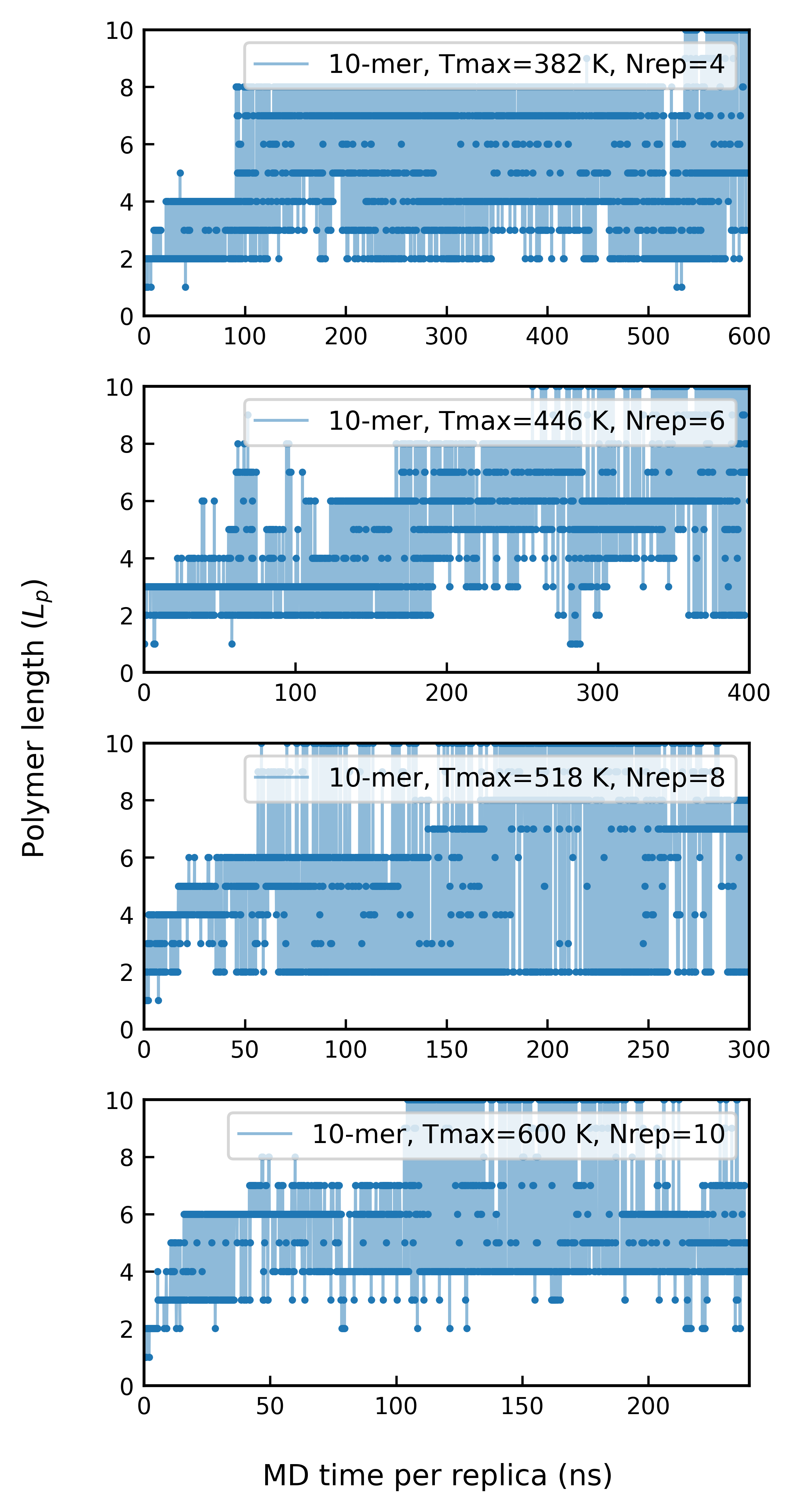}
\caption{\label{fig:grest10mer}
Evolution of the polymer length ($L_p$) for the 10-mer system
obtained from global REST simulations with
different values of the maximum replica temperature ($\Tmax$)
and the number of replicas ($\Nrep$). Elongated polymers are
efficiently produced when $\Tmax$ is slightly above
the aggregation-dispersion transition region.
}
\end{figure}

\textbf{Evolution of polymer length.}
\Fig{grest10mer} displays the evolution of $L_p$ for the target replica at 300 K.
The system rapidly forms a single aggregate within $\sim$50 ns,
followed by gradual polymer elongation driven by thermal rearrangements.
For $\Tmax = 382$ K, the polymer length increases to $L_p = 8$ within 100 ns,
but a significantly longer time ($\sim$500 ns) is required to reach $L_p = 10$.
In the following, we define $\tfull$ as the time at which a fully elongated polymer
first appears (namely, $\tfull \simeq 500$ ns in the above simulation).


The results further show that polymer growth becomes faster with increasing $\Tmax$.
Specifically, $\tfull$ decreases to 260 and 60 ns
for $\Tmax = 446$ K and 518 K, respectively,
suggesting enhanced monomer rearrangement at higher $\Tmax$.
However, when $\Tmax$ is increased further to 600 K,
$\tfull$ becomes longer again ($\sim$100 ns).
This slowdown is likely because higher-temperature replicas
increasingly sample molecularly dispersed configurations,
thereby significantly reducing the efficiency of replica traversal.

\begin{figure}[t]
  \centering
  \includegraphics[width=0.85\columnwidth]{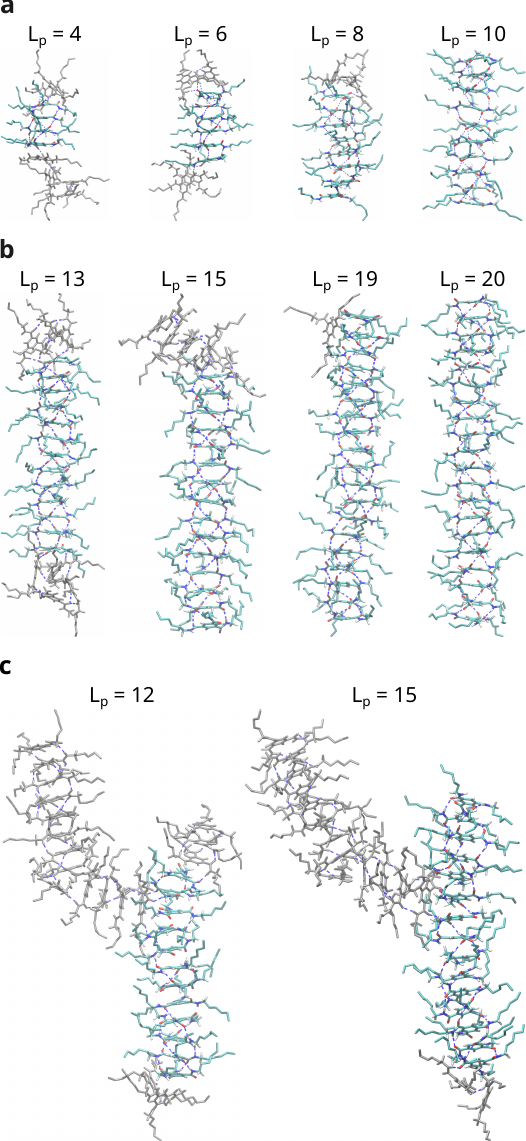}
\caption{\label{fig:grest_struct}
Typical structures obtained from global REST simulations of
(a) 10-mer, (b) 20-mer, and (c) 30-mer systems.
Disordered monomers are shown in gray to distinguish
them from elongated polymers of length $L_p$.
Blue dashed lines represent H-bonds between adjacent monomers.
The 10- and 20-mer simulations produce fully elongated polymers,
whereas the 30-mer simulation becomes trapped in branched configurations
with disordered terminal monomers.
}
\end{figure}

We note that $L_p$ spans a wide range of values, e.g., $L_p =$ 2--10 for $\Tmax = 518$ K.
Under the present simulation conditions
with the relatively high monomer concentration (0.038 M),
the statistical ensemble is not dominated by fully elongated polymers,
but instead contains a substantial population of more disordered states.
We summarize in \Fig{grest_struct} representative structures obtained from
global REST for different system sizes.
The results for the 10-mer system [panel (a)] show that
structures with low $L_p$ typically consist of short BTA stacks
with disordered terminal monomers.
On the other hand, high-$L_p$ structures frequently exhibit
neatly stacked BTAs with helical 2:1 H-bonding patterns,
in which two of the three amide hydrogens point in one direction
along the polymer axis, while the remaining hydrogen points
in the opposite direction.
A similar H-bonding pattern has been reported in previous studies.
\cite{Kulkarni2011,Cantekin2012,Bejagam2014,Arefi2017}
This pattern may arise from the balance between local H-bond interactions
and long-range electrostatic interactions associated with the macrodipole
along the polymer axis.


\begin{figure}[tb]
  \centering
  \includegraphics[width=0.9\columnwidth]{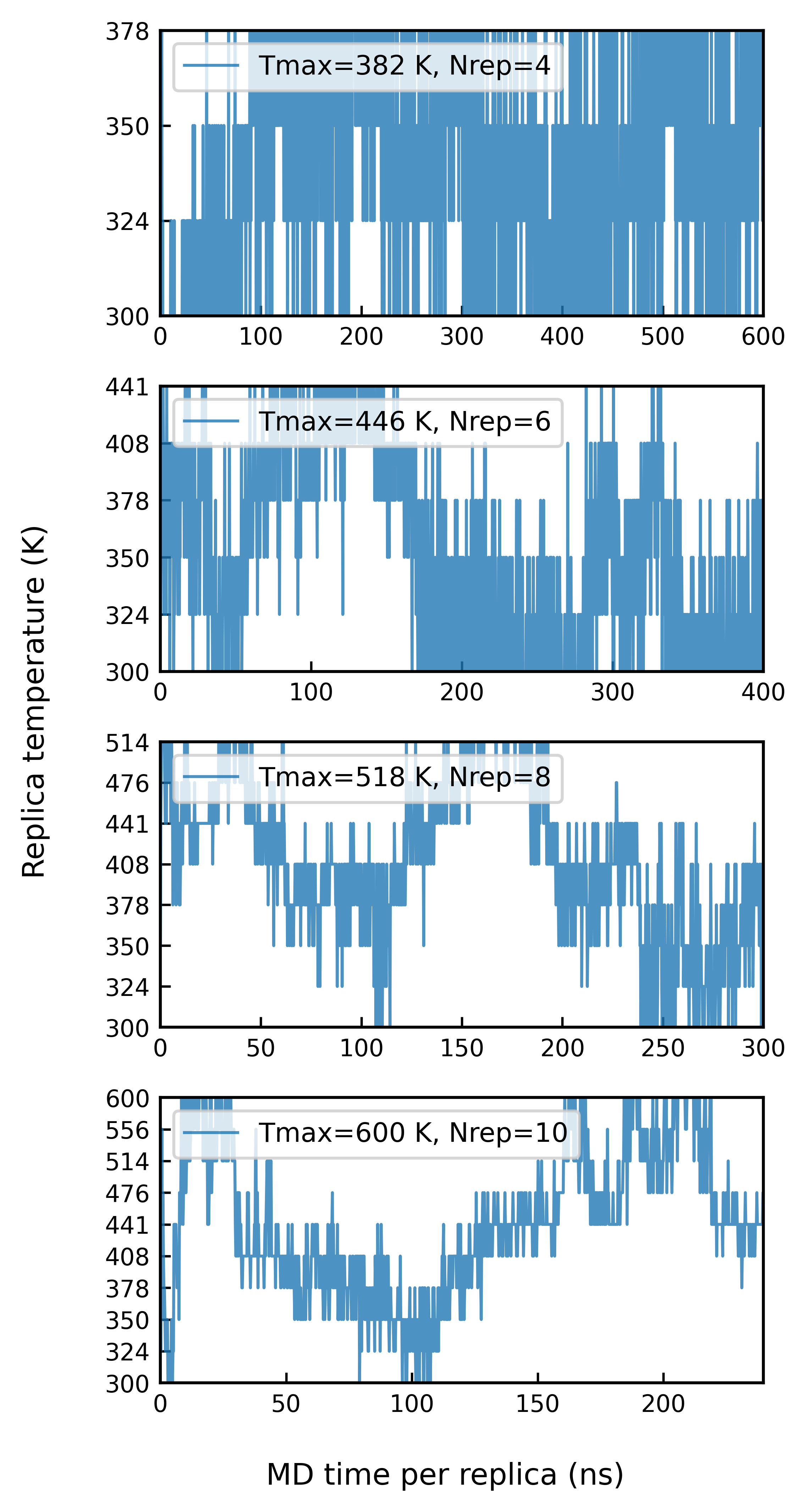}
\caption{\label{fig:grest10mer_reptemp}
Replica traversal in temperature space obtained from
global REST simulations of the 10-mer system.
When $\Tmax$ is above the transition-temperature region,
replica motion becomes slow and oscillatory between $T_0$ and $\Tmax$,
thereby reducing the efficiency of round trips.
}
\end{figure}


\textbf{Replica traversal in temperature space.}
\Fig{grest10mer_reptemp} displays the evolution of replica temperatures.
For relatively low $\Tmax$ (382 K),
replica traversal occurs efficiently between $T_0$ and $\Tmax$.
On the other hand, higher $\Tmax$ values (518 and 600 K)
produce slow oscillatory motion between $T_0$ and $\Tmax$,
indicating inefficient round trips in temperature space.
Similar behavior was also observed previously\cite{Arefi2017}
and is likely related to the aggregation-dispersion transition around 410--450 K.
Replica exchange methods are known to become inefficient
in systems exhibiting order-disorder phase transitions.\cite{Straub_GREM}
Although the present system is relatively small,
meaning that the transition does not correspond to
a true bulk phase transition, it nevertheless suffers from inefficient
replica traversal across the transition region.
Despite this limitation, global REST was still able to produce
fully elongated polymers, which were not observed
in long-time cMD simulations of the 10-mer system (\Table{cmd}).


\begin{figure}[tb]
  \centering
  \includegraphics[width=0.95\columnwidth]{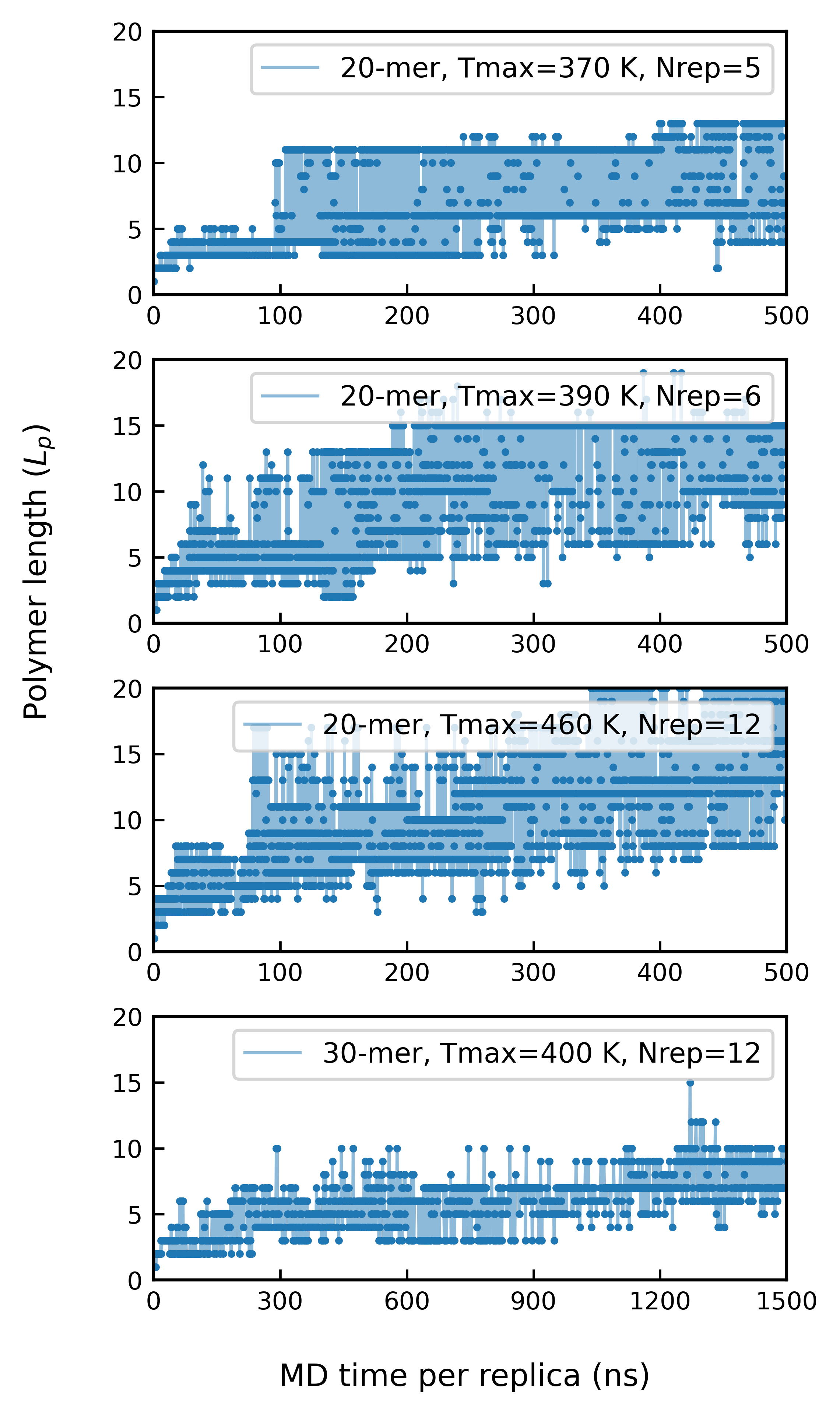}
\caption{\label{fig:grest20mer}
Evolution of the polymer length ($L_p$) obtained from
global REST simulations of the 20- and 30-mer systems.
The 20-mer simulation with sufficiently high $\Tmax$ (460 K)
produces fully elongated polymers, whereas the 30-mer system
fails to efficiently produce elongated polymers
within the accessible simulation time.
}
\end{figure}


\subsection{Results for larger systems}

\textbf{20-mer system.}
We next examine the performance of global REST for a larger system
consisting of 20 BTA and 4000 MCH molecules (\Table{grest}).
The monomer concentration is identical to that used
for the 10-mer system studied above (0.038 M).
\Fig{grest20mer} displays the evolution of $L_p$
for several different values of $\Tmax$.
The maximum value of $L_p$ for $\Tmax = 370$ and 390 K is
13 and 15 (with rare occurrences of $L_p = 19$), respectively.
Panel (b) in \Fig{grest_struct} shows that terminal monomers are
disordered in low-$L_p$ structures.
Further increasing $\Tmax$ to 460 K yields fully elongated polymers
with $L_p = 20$, although the latter simulation requires
substantially more replicas and consequently
a significantly higher computational cost.
These results are consistent with the previous observation that
efficient polymer elongation requires $\Tmax$
to be slightly above the transition-temperature region.\cite{Arefi2017}


We also note that the aggregation time in the REST simulation ($\sim$100 ns)
is significantly shorter than that in the cMD simulation ($>$1000 ns).
This is likely because REST also tempers solute-solvent
interactions\cite{Liu2005,Wang2011}, thereby accelerating monomer diffusion
via reduced solvent friction.
Similar behavior has been reported for the accelerated diffusion
of lipid molecules in previous REST studies.\cite{Smith2016}


\textbf{30-mer system.}
For comparison, we also performed global REST simulations
of a larger system containing 30 monomers.
Because fully elongated polymers can reach lengths of $\sim$100 \AA{},
we used a larger simulation box with a side length of 120 \AA{},
containing 8060 solvent molecules (corresponding to $\sim$58000 atoms).
To maintain an exchange probability of $\sim$20 \%
while keeping the simulations computationally feasible,
we used $\Nrep = 12$ with a relatively low value of $\Tmax$ (400 K),
close to the lower end of the transition-temperature region.
Under these conditions, each REST simulation was run for an extended time
(1500 ns per replica) to increase the probability of polymerization.

The simulation required a substantially longer time ($\sim$700 ns)
to produce a single large aggregate, likely due to the slow diffusion
of intermediate-sized aggregates. Even after aggregation,
the maximum value of $L_p$ remained 12 with rare occurrences of $L_p = 15$
throughout the simulation (\Fig{grest20mer}).
Typical structures with $L_p = 12$ and 15 are displayed
in \Fig{grest_struct}(c), showing branched BTA stacks
with disordered terminal monomers.
It would likely require both higher $\Tmax$ and longer simulations
to overcome these kinetically trapped states and achieve further
polymer elongation, although such conditions would substantially increase
the computational cost and further exacerbate replica traversal bottlenecks.


\begin{figure}[htb]
  \centering
  \includegraphics[width=0.65\columnwidth]{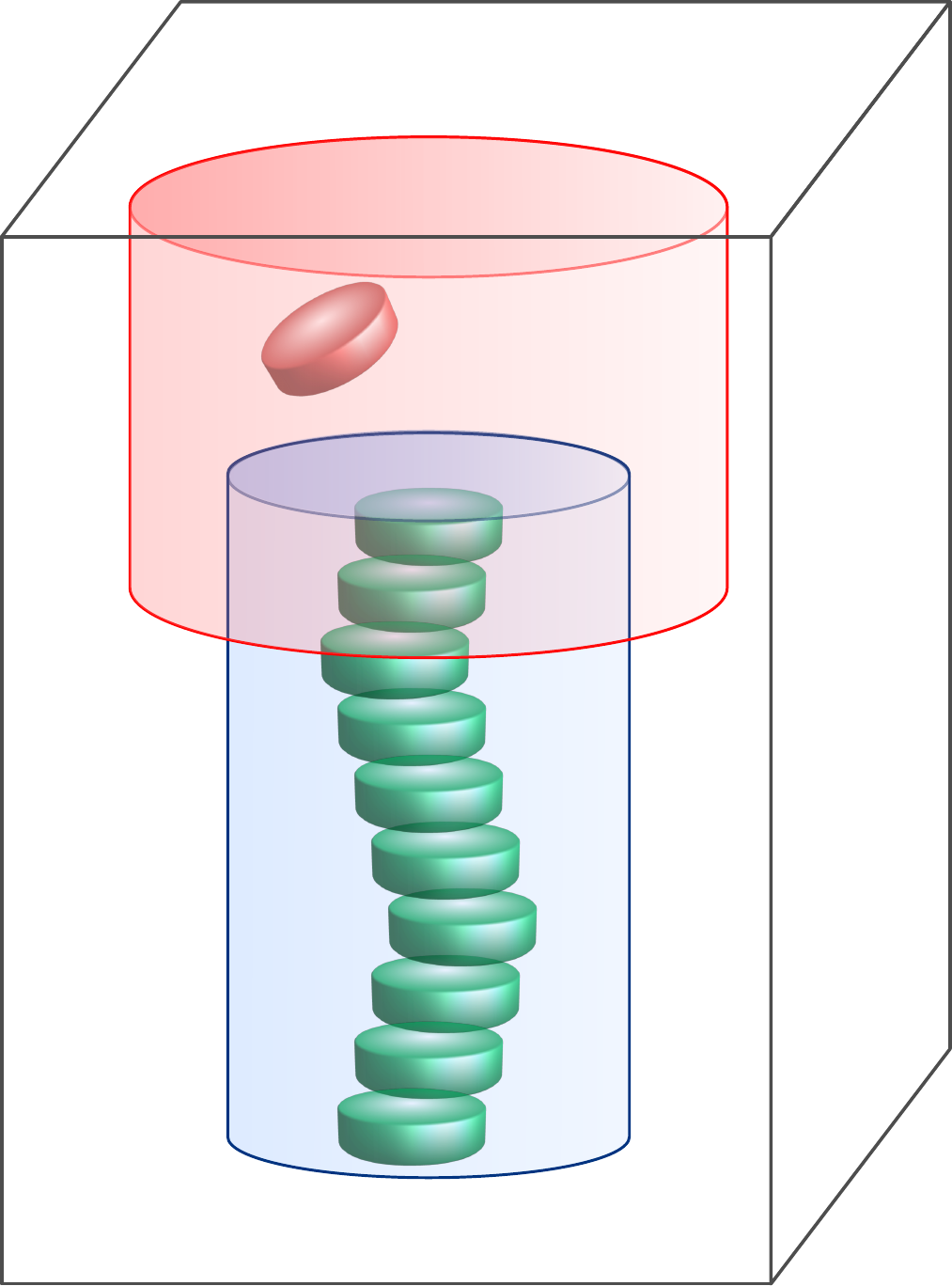}
\caption{\label{fig:seq_scheme}
Schematic illustration of the sequential REST approach.
Green disks represent existing BTA stacks restrained in
a cylindrical region (blue-shaded). The red disk indicates a newly added monomer
subjected to enhanced sampling for exploring favorable binding configurations.
The new monomer is also restrained in a separate cylindrical region (red-shaded)
near the polymer end.
}
\end{figure}


\section{Sequential REST} \label{sec:seqrest}  

\subsection{Basic idea}

In the previous section we have focused on the global REST simulations
that accelerate all the monomers simultaneously.
For the present BTA-C6 system the global REST was
able to generate fully elongated polymers up to 20 monomers.
This is encouraging because polymer structures
of this length could be used for studying, e.g., chemical and electronic
properties at the atomistic level.\cite{Botek2007}
On the other hand, the global REST simulations of the 30-mer system
were computationally expensive and failed to produce
ordered structures within the accessible simulation time.
Furthermore, global REST sampling is expected to become more demanding
for larger or more complex monomers because the number of
enhanced degrees of freedom increases.
These observations imply that global REST likely becomes
less effective for larger or more complex self-assembling systems.

To address these size-dependent sampling limitations,
we consider a construction-based approach that combines
sequential monomer addition with localized enhanced sampling.
Starting from a short seed, a new monomer is placed
near the polymer terminus and sampled to explore
favorable binding configurations.
Assuming a predominantly one-dimensional morphology,
we restrict the construction to terminal growth and
repeat this procedure to generate progressively
longer atomistic models of the supramolecular polymer.

A possible issue here is that a new monomer can be trapped in
metastable misbound states for a long time.
To avoid such kinetic trapping, we enhance the sampling of new monomers
by treating it as the ``solute'' region in REST while all the other molecules
as the ``solvent'' region. This helps significantly reduce the number of required
replicas because of the smaller size of the ``solute'' region,
thereby facilitating the use of high $\Tmax$ for more adequate sampling
of binding configurations.

In the following we study the above sequential approach
(hereafter referred to as sequential REST or seq-REST)
using the same BTA-C6 system and compare results with those obtained
from global REST. We also perform analogous monomer-addition simulations
based entirely on cMD and illustrate the benefit of local enhanced sampling
for new monomers. We note that the above approach is similar in spirit to
the application of REST to the problem of ligand binding,
\cite{Cole2014,Wang2013,Wang2015} in which REST is applied only
to ligand or drug molecules to identify favorable binding modes
on macromolecular surfaces. In the present study,
newly added monomers and existing aggregates play a role
similar to ligands and macromolecules, respectively.

\subsection{Stepwise elongation of BTA monomers}

\textbf{Computational details.}
We first summarize the computational details of seq-REST
(see \Fig{seq_scheme} for schematic illustration).
We first picked up short BTA stacks (5-mer) obtained
from the cMD simulation in \Sec{cmd}.
This aggregate structure was oriented in the $z$ direction and
placed at the center of a cubic simulation box of side length $L = 60$ \AA{}.
We then added a new monomer into the system
such that it was at least 6 \AA{} apart from the initial aggregate.
To restrict the sampling domain,
we further applied a weak restraint potential $V_r$ to the new monomer
such that $ V_r = V_c + V_z $, where $ V_c $ and $ V_z $ are
cylindrical and layer flat-bottom potentials, respectively.
This potential restricts the new monomer in a cylindrical region
with a radius of 10 \AA{} and a height of 15 \AA{} around the polymer axis.
(\Fig{seq_scheme}). The weak restraint potential allows
flexible sampling of monomer configurations near the polymer end,
while preventing complete dissociation into the bulk solvent.


Once the new monomer was added, the simulation box was filled with
solvent molecules, followed by short equilibrations under NVT and NPT conditions.
A REST simulation was then performed for 20 ns per replica
with only the new monomer subjected to enhanced sampling.
Because the corresponding ``solute'' region is rather small,
only 4 replicas were sufficient to span a wide range of replica temperatures,
$T_m =$ 300, 362, 436, and 526 K, while maintaining an exchange probability of 20--30 \%.
A favorable binding configuration of the new monomer was selected
from the REST trajectory by using a scoring function.
The selected configuration was used as the initial state
for the next step of monomer-addition simulations.
To accommodate a growing polymer, the simulation box was expanded
at each addition step such that the box size in the $z$ direction ($L_z$) was
$ 60 + 4 N $ \AA{} while $L_x$ and $L_y$ fixed at 60 \AA{},
where $N$ is the total number of added monomers.
We repeated the above procedure until a polymer of desired length was obtained.


\begin{figure*}[htb]
  \centering
  \includegraphics[width=0.95\textwidth]{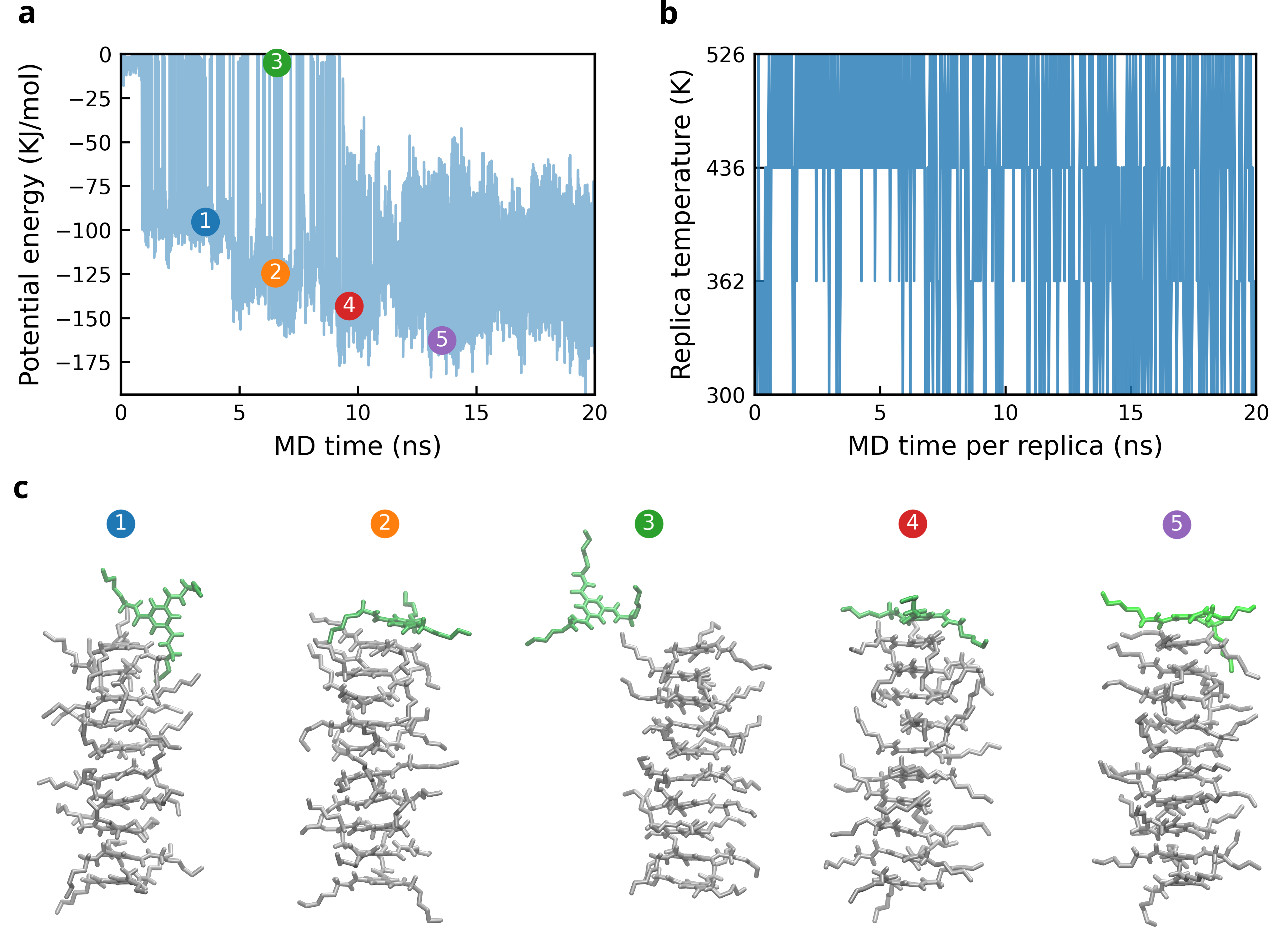}
\caption{\label{fig:seq_ene}
(a) Evolution of monomer-aggregate interaction energy $\Eint$ (kJ/mol) in \Eq{score}
during a representative addition step in sequential REST.
$\Eint$ is used to select favorable binding configurations of newly added monomers.
(b) Replica traversal in temperature space.
(c) Snapshots of the system corresponding to circles 1-5 in panel (a).
The newly added monomers (subjected to enhanced sampling)
are shown in green, while other monomers (evolved at ambient temperature) are in gray.
}
\end{figure*}



\textbf{Scoring function for monomer binding.}
To select favorable binding configurations of the monomer,
we need to define some scoring function ($\Escore$).
A rigorous thermodynamic ranking of alternative binding states
would require their relative free energies. Because such calculations
are computationally demanding, we instead use the monomer-aggregate
interaction energy as a simple scoring function for identifying
favorable configurations sampled during each local REST calculation.
%
For the present BTA system, in which intermolecular H-bonding
provides a major driving force for columnar assembly,
this interaction energy is expected to provide a useful approximate
measure of favorable monomer binding.
%
%
We thus define the scoring function as
\BE
\label{eq:score}
    \Escore \equiv \Eint( \mathrm{mon} \cdots \mathrm{agg} )
    .
\EE
Here, ``mon'' and ``agg'' denote the new monomer and existing
aggregate, respectively. The above $\Escore$
does not include solute-solvent and solvent-solvent interaction energies
because the role of MCH solvent is relatively minor for the present system.
The interaction energies within the existing aggregate are also excluded
because they introduce large fluctuations in the value of $\Escore$,
making the selection of binding configurations less straightforward.
With the above definition, we calculate $\Eint$ along
the REST trajectory obtained from each addition simulation
and select one configuration corresponding to the minimum of $\Eint$.
The latter is then used as the initial state for the next round
of addition simulations.


\Fig{seq_ene} (a) displays the evolution of $\Eint$ for the target replica
at a typical addition step. The new monomer is initially
dissociated from the aggregate, so that $\Eint$ starts from $\sim$0.
We note that the value of $\Eint$ evolves in a discontiguous manner
because the target replica frequently exchanges
its configuration with higher replicas and thus
the trajectory is discontiguous in time.
\Fig{seq_ene} (c) displays typical snapshots of the system
corresponding to the colored circles in \Fig{seq_ene} (a).
This figure shows that the monomer gradually relaxes
into more ordered stacking configurations.
The $\Eint$ decreases by a large amount when new H-bonds are formed
between the monomer and the existing aggregate.
From the REST trajectory thus generated,
we selected one configuration with the lowest value
of the scoring function (here equal to $\Eint$)
and used it as the initial state for the next addition simulation.

\textbf{Replica traversal in temperature space.}
\Fig{seq_ene} (b) displays the evolution of replica temperature.
As seen, the replica traversal occurs efficiently
with a large number of round trips even with
a sufficiently high $\Tmax $ (526 K).
This is in contrast to the global REST simulation
that exhibits a replica-trapping behavior
due to the aggregation-dispersion transition.
In global REST simulations,
the energy of the ``solute'' region (i.e., all monomers)
likely varies significantly when the replica traversal
occurs across the transition-temperature range.
This is because the breaking or formation
of a large number of H-bonds is involved in such
a change in the replica temperature.
On the other hand, the ``solute'' region in the seq-REST
simulation consists only of one monomer, which makes
the energy change upon replica traversal motions
significantly smaller than that in global REST simulations.
Furthermore, the weak restraint potential
helps avoid a long excursion of the monomer in the bulk solvent,
which facilitates rebinding of the monomer to the polymer end
upon replica traversal from high to low temperatures.
These differences likely make the diffusion in temperature space
significantly more efficient in seq-REST,
thus facilitating more extensive local sampling of binding configurations.

\begin{figure}[htb]
  \centering
  \includegraphics[width=0.9\columnwidth]{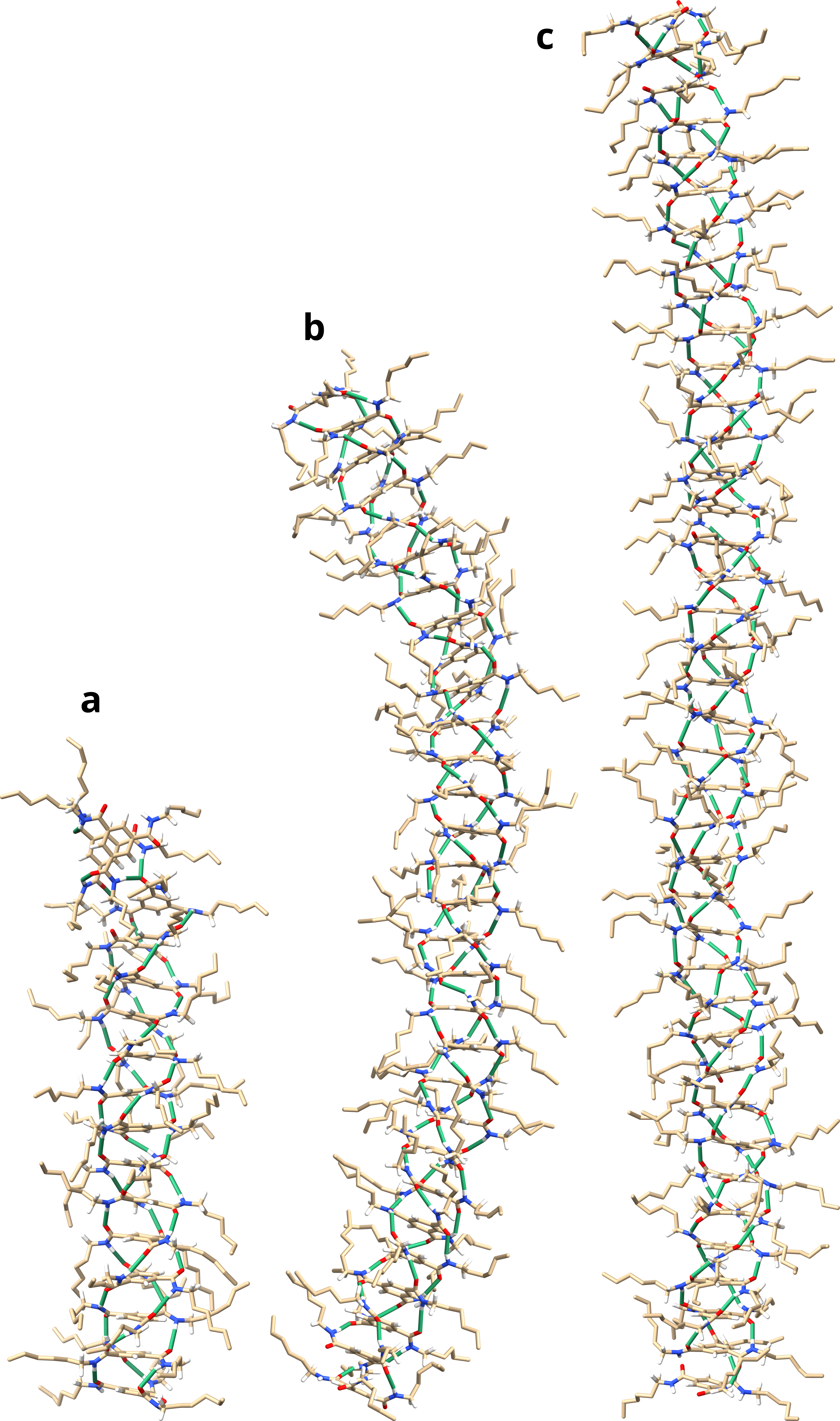}
\caption{\label{fig:seq_struct}
Typical structures obtained from sequential addition simulations.
(a) Result of addition simulations based entirely on cMD.
Two monomers were sequentially added to a pre-existing polymer,
but they were attached to the polymer end in a T-shaped manner,
thereby hindering further elongation.
(b,c) Result obtained from seq-REST for 20 and 30 monomers, respectively.
Addition simulations do not suffer from such kinetic trapping
and yield long polymers with characteristic H-bonding patterns.
}
\end{figure}

\textbf{Atomistic structures obtained from seq-REST.}
We repeated the sequential addition simulations
until the polymer was elongated up to 40 BTAs.
\Fig{seq_struct} (b) and (c) display the final polymer
structures consisting of 20 and 30 monomers.
We find that seq-REST produces neatly stacked polymers
with characteristic H-bonding patterns
similar to those observed in global REST simulations
(\Sec{grest}) and in previous studies.\cite{Bejagam2014,Arefi2017}
This result also suggests that for the present BTA system
the energy-based scoring function in \Eq{score}
was sufficient to generate the ordered stacks obtained here.
We repeated the entire addition steps several times and
confirmed that similar polymer structures can be obtained.

For comparison, we also performed addition simulations
based entirely on cMD, i.e., with no enhanced sampling for new monomers.
Specifically, we added two monomers sequentially to
a pre-existing polymer with $\Lp = 10$ (which
exerts an electric field of the macrodipole on new monomers)
and evolved the entire system via cMD.
\Fig{seq_struct} (a) displays typical binding configurations
observed in the addition simulations. This figure shows that
the new monomers are improperly bound to the polymer end
in a T-shaped manner, thereby hindering further elongation.
Although such defect structures could occur in natural polymers,\cite{Gasparotto2019}
we think that the misbound configurations in \Fig{seq_struct} (a) are
most likely due to the limited sampling capability of cMD.
This result highlights the benefit of local REST sampling
for avoiding kinetic trapping during sequential construction.


\section{Discussion} \label{sec:discuss}  

In this section we compare the advantages and limitations of
global and sequential REST and discuss possible improvements
for more complex systems.


\subsection{Global REST}

In \Sec{grest} we have studied the performance of standard REST
for accelerating all monomers in a global manner.
For systems up to 20 monomers, the global REST successfully produced
fully elongated polymers within a total MD time of several microseconds.
This is encouraging because global REST allows ordered assembly structures
to emerge from initially dispersed monomers without prescribing the target morphology.
On the other hand, we were not able to obtain elongated polymers
for the 30-mer system within our simulation time.
The above result suggests that the direct application of global REST
may be too demanding for studying larger or more complicated systems.

The performance of global REST is currently limited by several factors.
First, the number of replicas ($\Nrep$) needs to be increased with
the number of accelerated degrees of freedom.\cite{Fukunishi2002,Nymeyer2004}
While REST mitigates this problem by tempering only the ``solute'' molecules,
$\Nrep$ still needs to be increased with the number of those molecules and
the size of each molecule. Another factor that limits the performance
is order-disorder transitions inherent in self-assembling systems.
To sample configurations extensively,
one needs to set the maximum replica temperature ($\Tmax$)
sufficiently high so that the system can break intermonomer H-bonds.
However, the use of high $\Tmax$ also allows monomers to dissociate
from each other, resulting in repeated aggregation-dispersion transitions.
Those transitions are likely responsible for
the very inefficient replica traversal observed in \Fig{grest10mer_reptemp}
and in our previous study.\cite{Arefi2017}
A similar issue of replica trapping behaviors
is also known, e.g., for the folding transition of proteins\cite{Garcia_REMD_review}
and magnetization of spin systems.\cite{Straub_GREM}

The issue of replica trapping may be addressed by several approaches.
For example, it may be beneficial to apply a weak restraint potential that
prevents monomers from dissociating into the bulk solvent.
While it is important for monomers to be able to break H-bonds
and correct for disordered configurations,
it is clearly inefficient for monomers to dissociate completely
and spend too long times in the bulk solvent.
A suitable restraint potential may help avoid such complete
dissociation.
Another approach may be to use enhanced-sampling techniques
that are more robust against order-disorder transitions
(e.g., the generalized replica-exchange method).\cite{Straub_GREM}
It should be noted, however, that even with the above approaches
the scalability limitations of global REST likely persist,
i.e., the required number of replicas increases with system size.
This seems to be one computational bottleneck that hampers routine applications
of global REST to more complex systems.

\subsection{Sequential REST}

Based on the above observations, we have explored an alternative
approach for polymer structures by performing sequential addition of monomers
and accelerating their binding motions via REST.
The main advantage here is that the number of replicas depends
primarily on the size of the locally tempered region,
which in the present implementation consists of a single incoming monomer.
This allowed us to use only four replicas at each addition step.
The above approach has been tested with the same BTA-C6 system and was
shown to generate neatly stacked BTAs with characteristic H-bonding patterns.


Here we make a few remarks on the properties of the sequential approach:

(1) At each step of monomer addition, only the new monomer
is accelerated via REST to explore favorable binding configurations.
This does not mean that the other existing monomers are fixed at some
configuration, but they are also evolved as the ``solvent'' region of the REST scheme.
Thus, the existing monomers (particularly those close
to the polymer end) can also make conformational changes
upon binding of the new monomer (within the capability of cMD).

(2) The individual addition steps are not equivalent to each other.
Specifically, we recall that a growing BTA polymer develops
a macrodipole along the polymer axis.\cite{Kulkarni2011,BalaReview2018}
At each addition step a newly added monomer experiences
a different macrodipole created by the existing polymer.
Thus, individual addition steps are not equivalent:
the local environment experienced by an incoming monomer changes
as the aggregate grows.


(3) We emphasize that seq-REST is a type of structural modeling method
rather than a dynamical simulation of polymer elongation.
This distinction is important because the actual polymerization process
may be quite different from sequential monomer addition.\cite{Bochicchio2017a}
For example, dispersed monomers may first form a set of short polymer fragments,
followed by their association into long polymers.
Polymerization processes may also depend strongly
on monomer concentration.\cite{Bochicchio2017a}
The purpose of the present method is not to reproduce
such dynamical information, but to generate atomistic structural models
under an assumed one-dimensional morphology and terminal-growth construction protocol.


\textbf{Including solvent effects into the scoring function.}
As a scoring function ($\Escore$) for favorable binding configurations,
we have used the gas-phase interaction energy ($\Eint$ in \Eq{score})
between the new monomer and the existing aggregate.
This is because the polymerization of the present system is mainly driven
by multiple H-bonds between BTA molecules, while the solvent effect of
nonpolar MCH molecules is rather small. On the other hand,
for systems in polar solvents (particularly in water) the solvent effect
can affect the self-assembled structure rather strongly, and hence
it is important to include the solvent effect somehow into the scoring function.
(We note in passing that the seq-REST simulation is performed
with an explicit solvent model, so that the REST trajectory itself
already includes solvent effects. The question here is how to choose
thermodynamically favorable binding states along the REST trajectory.)

Ideally, one could calculate the free energy of the total system
$\Gtot$ for different configurations and compare their values
to rank alternative binding states. However, the rigorous calculation of $\Gtot$ is
usually too demanding for large assembly systems.
In this case it may be useful to follow the popular approach used
in solution-phase chemistry, i.e., to represent $\Gtot$
as the sum of the gas-phase interaction energy $\Eint$ and an
approximate solvation free energy $\Gsolv$, such that
\BE
  \Gtot \simeq \Eint + \Gsolv.
\EE
The $\Gsolv$ may be evaluated efficiently with a solvation model
for nano-sized systems (for example, see \Ref{Borgis_MDFT_JCP21}).
One could then evaluate $\Gsolv$ along the REST trajectory
and choose the lowest free-energy configuration for use in the next addition step.
Depending on systems, the above type of approach may be required to
adequately describe solvent effects on polymer elongation.


\textbf{Treatment of more complex systems.}
In the above sections we have used a relatively simple BTA system
in nonpolar solvent for testing the seq-REST approach.
On the other hand, supramolecular polymers
of experimental interest often consist of more complex monomers
with long side chains; for example, water-soluble BTAs often possess
quite long amphiphilic side chains for increasing the solubility in water.
\cite{Pavan2016water}
One can readily expect that the presence of such side chains
will make seq-REST more challenging because of the possible entanglement
of long side chains among monomers. To address this issue,
it may be useful to accelerate not only a newly added monomer but also
several monomers near the polymer end, such that their
collective motions will be enhanced simultaneously.
The corresponding seq-REST simulation may still be feasible
because of the relatively low computational demand as compared to global REST.
Testing such an approach for more complex systems
remains a subject of future study.

\section{Conclusion}  \label{sec:concl}  

In this work, we explored a construction-based strategy for atomistic modeling
of supramolecular polymers by combining sequential monomer addition
with localized enhanced sampling. Benchmark calculations with
conventional MD and global REST showed that sampling becomes
increasingly difficult as the assembling system grows.
Although global REST produced fully elongated BTA polymers
for systems up to 20 monomers, the 30-mer system remained
trapped in partially ordered structures within the accessible simulation time,
and replica traversal became increasingly hindered by
aggregation-dispersion behavior. These observations motivated
a sequential strategy in which REST is applied only to
an incoming monomer near the growing polymer terminus.

For the BTA-C6 system studied here, sequential REST enabled
continued construction of ordered stacks up to 40 monomers
while maintaining characteristic hydrogen-bonding motifs
consistent with those observed in global REST calculations.
In contrast, a corresponding sequential cMD test became kinetically
trapped and failed to sustain ordered elongation,
illustrating the role of localized enhanced sampling
in facilitating monomer rearrangement during stepwise growth.
The sequential approach should not be interpreted as a simulation
of spontaneous polymerization or as a procedure for obtaining
the complete equilibrium ensemble. It assumes predominantly
one-dimensional terminal growth, and the resulting structures
can depend on the configurations selected at preceding
construction steps. Within these limitations,
sequential REST provides a practical construction-based route
for generating atomistically resolved models of supramolecular
polymers while avoiding the need for global enhanced sampling
of the entire growing assembly.


\section*{Supplementary Material}   
Further details of the simulation protocols and molecular mechanical
force fields are presented.


\begin{acknowledgments}   
This work was supported by JSPS KAKENHI Grant Number JP19K05385.
\end{acknowledgments}


\section*{References}   

\bibliography{ref}


\end{document}